%Paper: astro-ph/9309055
%From: "T. Buchert " <TOB@IBMA.ipp-garching.mpg.de>
%Date: Thu, 30 Sep 93 12:27:19 MET

\magnification=1200
\tolerance=1400
\overfullrule=0pt
\baselineskip=15pt
\font\rmc=cmr9
\font\rmb=cmr9 scaled \magstep 1
\font\rma=cmr9 scaled \magstep 2
\font\rmm=cmr9 scaled \magstep 3
\def\ueber#1#2{{\setbox0=\hbox{$#1$}%
  \setbox1=\hbox to\wd0{\hss$\scriptscriptstyle #2$\hss}%
  \offinterlineskip
  \vbox{\box1\kern0.4mm\box0}}{}}

\def\R{\rm I\kern-.18em R}
\def\etal{{\it et al. }}
\topskip 6 true cm
\pageno=0

\centerline{\rmm LAGRANGIAN THEORY OF GRAVITATIONAL INSTABILITY}
\smallskip
\centerline{\rmm OF FRIEDMAN-LEMAITRE COSMOLOGIES -}
\bigskip
\centerline{\rmm GENERIC THIRD ORDER MODEL}
\medskip
\centerline{\rmm FOR NON-LINEAR CLUSTERING}
\bigskip\bigskip
\centerline{\rmm by}
\bigskip\bigskip
\centerline{\rmm Thomas Buchert}
\vskip 2 true cm
\centerline{\rma Max-Planck-Institut f{\"u}r Astrophysik}
\smallskip
\centerline{\rma Karl-Schwarzschild-Str. 1}
\smallskip
\centerline{\rma D-85740 Garching, Munich}
\smallskip
\centerline{\rma G e r m a n y}
\bigskip
\vskip 3 true cm
\centerline{\rma {\it M.N.R.A.S.}, accepted}
\vfill
\eject

\centerline{\rmm Lagrangian theory of gravitational instability}

\centerline{\rmm of Friedman-Lema\^\i tre cosmologies -}

\centerline{\rmm generic third order model for non-linear clustering}
\bigskip\bigskip
\centerline{\rma by}
\bigskip\bigskip
\centerline{\rma Thomas Buchert}
\vskip 1.3 true cm
\noindent
{\rmb
{\narrower
{\rma Summary:}
The Lagrangian perturbation theory on Friedman-Lema\^\i tre
cosmologies
investigated and solved
up to the second order in earlier papers
(Buchert 1992, Buchert \& Ehlers 1993) is evaluated up to the
third order. On its basis a model for non-linear clustering
applicable to the modeling of large-scale
structure in the Universe
for generic initial conditions is formulated.
A truncated model is proposed which represents
the ``main body'' of the perturbation
sequence in the early non-linear regime by
neglecting all gravitational sources which describe
interaction of the perturbations.
However, I also give the irrotational
solutions generated by the interaction
terms to the third order,
which induce vorticity in Lagrangian space.
The consequences and applicability of the solutions are put into
perspective.
In particular, the model presented
enables the study of previrialization effects in gravitational
clustering and
the onset of non-dissipative
gravitational turbulence within the cluster environment.

}}

\vfill\eject
\topskip= -6 true cm

\noindent{\rmm 1. Introduction}
\bigskip\bigskip
{\rmb\noindent
For a long period of research in cosmology inhomogeneities in the
Universe have been modeled on the basis of a perturbative approach
exploiting the instability of the standard
cosmologies of Friedman-Lema\^\i tre type against perturbations
of the density and the velocity field. This approach
is {\it Eulerian}, i.e., the perturbations are evaluated
as a function of Eulerian coordinates (see, e.g., Peebles 1980). The
limitations of this
approach have been widely recognized, since it relies on the smallness
of physical densities which is inappropriate for the modeling of the
high density excesses observed in the Universe.
Zel'dovich (1970, 1973) has realized this situation and has proposed
an approximate extrapolation of the linear perturbation solution into
the non-linear regime.

\smallskip
In recent papers
the {\it Lagrangian} theory of gravitational instability of
cosmologies of
Friedman-Lema\^\i tre type for preasureless (``dust'') matter
has been investigated and solved up to the second order in
the deviations from homogeneity (Buchert 1992, Buchert \& Ehlers 1993,
henceforth abbreviated by B92 and BE93). This theory does not rely
on the smallness of the density of the
inhomogeneities, only the deviations
of the particle
trajectories from the homogeneous Hubble flow are treated
perturbatively. This is possible, since the field of trajectories
$\vec f (\vec X,t)$
is the only dynamical variable in the Lagrangian picture.
Interestingly, the widely applied
``Zel'dovich approximation"
for modeling the formation of large-scale structure
in the Universe was found to be contained in a
subclass of first order irrotational perturbation solutions
in this theory (B92). The first order solutions have been analyzed
in an earlier paper (Buchert 1989) in which
they are shown to
provide exact three-dimensional
solutions in the case of {\it locally} one-dimensional
motion, i.e., when two eigenvalues of
the peculiar-velocity gradient vanish along trajectories of fluid
elements. The general plane-symmetric case ({\it globally}
one-dimensional motion) constitutes a subclass of these solutions.

\smallskip
It should be emphasized that solutions of the Euler-Poisson system,
exact
or in a perturbative sense, will depend
{\it non-locally} on the initial conditions
for the inhomogeneities. In general, they have to be constructed
by solving elliptic boundary value problems, as will be discussed
in full detail. Despite this, Zel'dovich's approximation
is local in this sense.
It can be made rigorous that the first order perturbation
solutions can be made local without loss of generality (B92). This
will be recalled in APPENDIX B. The problem of uniqueness in
general perturbation solutions will be considered in a separate
paper (Ehlers \& Buchert 1993).

\smallskip
The success of Zel'dovich's model as an approximation
for irrotational self-gravitating ``dust'' flows in the early
non-linear regime is commonly appreciated.
Its range of applicability
can be roughly limited to the epoch shortly after the first
shell-crossing singularities in the flow develop,
and provides an excellent
approximation for the density field down to the non-linearity scale
(i.e., where the r.m.s. deviations of the density
from homogeneity exceed unity)
in comparison with numerical N-body simulations (Coles \etal 1993).
In contrast to other analytical models for the formation of large-scale
structure, the Lagrangian perturbation theory offers a systematic
and explicit
way to extend the range of validity of Zel'dovich's model
for generic initial conditions
(Buchert 1993).
While the first order approximation chiefly covers the (up to the epoch
of shell-crossing dominating) kinematical aspects of the structure
formation process, the second order approximation firstly involves
the tidal action of the gravitational field.
The collapse process is significantly accelerated by this action.
Also, tidal forces constitute the essence of so-called
previrialization effects in
gravitational clustering (Peebles 1990, BE93).

\smallskip
Zel'dovich's model has shortcomings after the epoch when shell-crossing
singularities develop, since shells of matter are predicted to cross
freely, and the kinetic energy of particles at the caustic exceeds the
gravitational potential energy. As a result no potential
well is formed in contradiction with numerical simulations.
The advantage of high-spatial resolution N-body simulations
is obvious especially with respect to the modeling of
multi-stream systems
which develop after shell-crossing. Analytically, this modeling is not
straightforward, since the particles are not treated as interacting
bodies. Rather, the particles are viewed as tracers of a flow, which
generally looses uniqueness as shells of matter overlap.
However, while the Eulerian representation
of any solution breaks down as soon as singularities in the
density field (caustics) form, the Lagrangian representation allows
to follow the trajectories $\vec f$ of fluid elements
across
caustics. The solution $\vec f$ remains regular, only the transformation
from Eulerian to Lagrangian space is singular. Non-uniqueness of the
flow is only realized in Eulerian space, where a patch of matter
can originate from different Lagrangian `particles', we say that the
flow consists of several streams in such a region.
Therefore, analytical Lagrangian approximations are in principle capable
to model fully developed non-linear situations including multiple
shell-crossings.

\smallskip
The phenomenology of a self-gravitating multi-stream system
is complex, the number of streams is systematically increasing
in time. Numerical simulations demonstrate that a hierarchy of
singularities forms (e.g. Doroshkevich \etal 1980), shells do not
cross freely without
limit as e.g. predicted by Zel'dovich's model, rather the gravitational
action of the firstly formed
three-stream system (called `pancakes' in the cosmological context)
forces the inner fluid elements
to recollapse and participate in a second shock structure and so on.
This way a five-, seven-, nine-stream system etc. develops.
A potential well is established leading to a self-trapped
quasi-stationary structure around the firstly developed singularities.
The morphogenesis of such a non-dissipative structure (e.g. a cluster
of galaxies in the present context) is similar to that of a flower:
the singular density peaks split up into shock structures
which subsequently emerge from and move away from the center.
Apparently, a cluster is composed of expanding shells,
while permanently new shells are created in
the center. A nested moving system of singularities is formed on
smaller and smaller spatial scales.
In real terms the situation is more complicated:
in addition higher-order singularities (such as the `breaking' of a
pancake boundary) increase the number of streams
as was originally discussed for Zel'dovich-Arnol'd pancakes
by virtue of analyzing a
numerical simulation in two spatial dimensions
(Arnol'd \etal 1982). The resolution of
higher-order singularities can also be appreciated in high-resolution
plots of
analytical mappings such as the ``Zel'dovich approximation'' (in 2D:
Buchert 1989, in 3D: Buchert \& Bartelmann 1991).

\smallskip
Analytical models have also been used to analyze self-gravitating
multi-stream systems: Approximating a generic density peak locally by
a triaxial ellipsoid, a model for the
evolution of this peak has been formulated by Gurevich and Zybin
(1988a,b) by performing a transformation of the known spherically
symmetric solution.
They show that the density distribution around the collapsed peak
can be represented by a
scaling law with a power of $\approx -1.8$ in this model.
This law is found to hold on a wide range of spatial scales from the
scale of clusters down to the scale of typical star distances.
Also Moutarde \etal (1991) found a similar scaling behavior of the
density field in the early non-linear regime of the collapse.
The correspondence to the scaling behavior of observed galaxy
concentrations as measured by the two-point correlation function is
striking and suggests that multi-stream hierarchies in the density
field of dark matter might
be responsible for the clustering properties
of galaxies (Gurevich \& Zybin 1988a, Berezinsky \etal 1992).
However, this conclusion depends on the assumption that
the dark matter dominates the matter density down to these scales
(Gurevich \& Zybin 1990).
It is likely that the fraction of baryonic matter depends on scale
and position which implies that the influence of
hydrodynamic or radiative effects of the baryonic
component might alter this picture.
Also, it is not shown that a generic collapse occurs at the peaks of
the {\it initial} density field. It is likely that, in general,
a collapse occurs closer to the trajectory of the maximum of the largest
eigenvalue function of the initial displacement tensor (Shandarin, priv.
comm.).
Gurevich and Zybin call this hierarchical structure
``non-dissipative turbulence''. Its existence is related to the early
conjecture by Mandelbrot (1976) that singularities in self-gravitating
flows might form a fractal, although they rather display a
multi-fractal scaling (Mart\'\i nez 1991, Jones 1993).
Indeed, a self-similarity of caustic patterns can be appreciated
(see BE93) in accord with the existence of self-similar solutions
(Filmore \& Goldreich 1984, Bertschinger 1985).
Moreover, in developed gravitational turbulence the density scaling
law seems to be generic, i.e., it seems not to depend on the initial
fluctuation spectrum according to Gurevich and Zybin's model.
However, it appears, at least with the amount of non-linearity one
can resolve, that the impact of large-scale power will destroy
the convergence to this local power law form, thus
retaining the memory from the initial conditions,
(compare the final power spectra for different intial conditions in
Melott \& Shandarin 1990).

\smallskip
The value of Lagrangian perturbation solutions has to be tested against
the phenomenology of ``non-dissipative turbulence''.
Indeed it has been demonstrated that the second order solutions
do describe the onset of such a hierarchy. They predict a second
shell-crossing singularity within Zel'dovich-Arnol'd pancakes.
A second bifurcation branch appears on the critical manifold of the
flow
in Lagrangian space (BE93). This prediction suggests that m-th order
Lagrangian perturbations will transform pancakes into
2m+1-stream systems in the coarse of time. The stage until when
the
perturbation solutions are valid could be estimated by the time when the
shell-crossing singularities of the corresponding order appear.

\smallskip
In the present paper, I evaluate the third order perturbation
solutions. In this line the need for higher order solutions
for the purpose of modeling large-scale structure should be
questioned. Certainly, it will not be adequate to derive higher
orders unless we can expect to obtain an appropriate model
for large-scale structure which remains sufficiently simple
to handle with respect to applications, apart from the fact that
the derivation of higher order perturbation solutions for generic
initial conditions is cumbersome.
As an argument to go to the third order I consider the structure of
the equations to be solved: they are cubic in the basic
dynamical variable (see B92, APPENDIX), so we can expect that a third
order
solution will cover the main effects of the perturbation sequence in the
early non-linear regime.
We first derive the longitudinal perturbations, i.e.,
for the case where the perturbations admit a potential in Lagrangian
space. In contrast to opinions stated in the literature
(Moutarde \etal 1991, Bouchet \etal 1992, Lachi\`eze-Rey 1993),
this restricts the generality of the solutions
as will be discussed in detail.
The restriction vanishes if {\it interaction} of perturbations
(among the first and second order perturbations here) is neglected,
an assumption which will define a class of third order models
which we consider the ``main body`' of the perturbation sequence.
Since, in general, the interaction terms generate vorticity in
Lagrangian space
(even for {\it irrotational} flows considered throughout this paper),
there will be a
transverse part of the third order solution, which I shall derive.

\smallskip
As an example the
approximation is evaluated on an Einstein-de Sitter
background for initial
conditions which correspond to Zel'dovich's model, and the result
is expressed in terms of the initial
conditions for the peculiar-velocity potential, or the
peculiar-gravitational potential, respectively.
With this specification of the initial conditions (discussed in
section {\bf 3}) we have assumed a functional relationship between
the peculiar-velocity potential and the peculiar-gravitational potential
(here: proportionality). Relaxing this relationship will also induce
vorticity in Lagrangian space already at the second order level
(see BE93).

}

\vfill\eject

\noindent{\rmm 2. The perturbation equations in Lagrangian form}
\bigskip\medskip
{\rmb\noindent
Let us briefly summarize the formal consequences of a Lagrangian
description of self-gravitating flows in Newton's theory.
We introduce integral curves
$\vec x = \vec f(\vec X,t)$ of the velocity field
$\vec v(\vec x,t)$:
$$\eqalignno{& {d \vec f \over dt} = \vec v (\vec f, t) \;\; , \;\;
\vec f(\vec X, t_o) = :\vec X \;\; . &(1)\cr}
$$
These curves are labelled by the Lagrangian
coordinates $X_i \,$; $x_i$ are non-rotating Eulerian coordinates. We
can express all Eulerian fields
in terms of the field of trajectories
$\vec f$ as was explained in Buchert (1989), B92 and BE93.
We denote the determinant of the deformation tensor $(f_{i,k})$ by $J$,
where the comma denotes
partial differentiation with respect to the Lagrangian
coordinates, a dot will denote Lagrangian time derivative
${d \over dt} := \partial_t \vert_x + \vec v \cdot
\nabla_x = \partial_t \vert_X$; comma and
dot commute.
Recall that mass conservation is guaranteed in the Lagrangian
representation irrespective of
{\it any} equations the trajectories $\vec f$ might obey. The
Euler-Poisson system for ``dust'' matter
can be cast into a set of four evolution equations
for the single dynamical variable $\vec f$ (Buchert \& G\"otz 1987,
Buchert 1989):
$$
\eqalignno{&\epsilon_{pq \lbrack j} {\partial ({\ddot f}_{i
\rbrack},
f_p,f_q) \over \partial(X_1,X_2,X_3)} = 0 \;\;\;,\; i \ne j \;\;,\;\;
&(2a,b,c) \cr
\sum_{a,b,c}\;{1 \over 2}&\epsilon_{abc}\;{\partial({\ddot f}_a,f_b,f_c)
\over \partial (X_1,X_2,X_3)} \; - \Lambda \,J
\;\, = \;\; - 4 \pi G \; \ueber{\rho}{o}(\vec X)\;\;;\;\;
\ueber{\rho}{o}(\vec X) > 0 \;\;. &(2d,e) \cr}
$$
(indices run from 1 to 3, if not
otherwise explicitly stated; henceforth,
$\nabla_0$ denotes the nabla operator with respect to the Lagrangian
frame which commutes with the dot).

\medskip
We proceed to the evaluation of the perturbation equations. (The
reader may consult earlier papers for more details.)
We first make the following perturbation ansatz for {\it longitudinal}
perturbations superposed on an isotropic homogeneous
deformation:
$$
\vec f = a(t)\vec X + \vec p \;\;;\;\;\vec p = \varepsilon \nabla_0
\psi^{(1)} + \varepsilon^2 \nabla_0 \psi^{(2)} + \varepsilon^3
\nabla_0 \psi^{(3)} \;\;.\eqno(3a)
$$
The parameter $\varepsilon$ is supposed to be small and dimensionless.
It can be considered as the amplitude of the initial perturbation
field.
We formally split the initial density accordingly:
$$
\ueber{\rho}{o} = \ueber{\rho_H}{o} + \varepsilon
\delta\ueber{\rho}{o}^{(1)}
+ \varepsilon^2 \delta\ueber{\rho}{o}^{(2)} + \varepsilon^3
\delta\ueber{\rho}{o}^{(3)}
\;\;.\eqno(3b)
$$
However, we can choose initial data at our convenience, so we put
$$
\delta\ueber{\rho}{o}^{(1)}:= \delta\ueber{\rho}{o} =: \ueber{\rho}{o}_H
\ueber{\delta}{o} \;\;,\;\;
\delta\ueber{\rho}{o}^{(2)}:=0 \;\;,\;\;
\delta\ueber{\rho}{o}^{(3)}:=0 \;\;,\eqno(3c)
$$
where $\delta\ueber{\rho}{o}$ denotes the total initial density
perturbation, $\ueber{\delta}{o}$ the initial density contrast.
This formality is adequate, since the density needs not
be perturbed in the Lagrangian framework.
\smallskip
\noindent
In what follows,
$I(\psi_{,i,k}) = tr(\psi_{,i,k}) = \Delta_0 \psi$,
$II(\psi_{,i,k}) = {1 \over 2} \lbrack(tr(\psi_{,i,k}))^2 -
tr((\psi_{,i,k})^2)\rbrack$ and $III(\psi_{,i,k})=\det(\psi_{,i,k})$
denote the three principal scalar invariants of the tensor
$(\psi_{,i,k})$.
For the derivation of the perturbation solutions we shall first
concentrate
on the source equation (2d) and consider the conditions
(2a,b,c) as constraint equations, which are checked after the
solution is obtained;
the constraint equations and the resulting constraints on
initial conditions for the longitudinal third order approximation are
given in APPENDIX A. We shall then construct the transverse
part which removes the constraints obtained for the pure longitudinal
solution.

\medskip\noindent
Inserting the ansatz (3) into equation (2d), we obtain
the following set of equations to be solved:
\smallskip
$$
\varepsilon^0 \Bigl\lbrace
- 4 \pi G \ueber{\rho}{o}_H = 3 {\ddot a} a^2 - a^3 \Lambda
\Bigr\rbrace
= 0 \eqno(4a)
$$
$$
\varepsilon^1 \Bigl\lbrace - 4 \pi G \delta\ueber{\rho}{o}^{(1)} =
\Bigl\lbrack (2 {\ddot a} a - a^2 \Lambda)
+ a^2 \; {d^2 \over dt^2} \Bigr\rbrack \Delta_0 {\psi}^{(1)}
\Bigr\rbrace
= 0 \eqno(4b)
$$
$$
\varepsilon^2 \Bigl\lbrace - 4 \pi G \delta\ueber{\rho}{o}^{(2)} =
\Bigl\lbrack (2 {\ddot a} a - a^2 \Lambda)
+ a^2 \; {d^2 \over dt^2} \Bigr\rbrack \Delta_0 {\psi}^{(2)}
\;\;\;\;\;\;\;\;\;
$$
$$
+ ({\ddot a} - a \Lambda) II(\psi^{(1)}_{,i,k})
+ a \sum_{a,b,c} \epsilon_{abc}\;{\partial({\ddot
\psi}^{(1)}_{,a},\psi^{(1)}_{,b},X_c) \over \partial (X_1,X_2,X_3)}
\Bigr\rbrace = 0   \eqno(4c)
$$
$$
\varepsilon^3 \Bigl\lbrace - 4 \pi G \delta\ueber{\rho}{o}^{(3)} =
\Bigl\lbrack (2 {\ddot a} a - a^2 \Lambda)
+ a^2 \; {d^2 \over dt^2} \Bigr\rbrack \Delta_0 {\psi}^{(3)}
\;\;\;\;\;\;\;\;
$$
$$
+ ({\ddot a} - a \Lambda) \sum_{a,b,c} \epsilon_{abc}\;
{\partial(\psi^{(2)}_{,a},\psi^{(1)}_{,b},X_c)
\over \partial (X_1,X_2,X_3)}
\;+\; a \sum_{a,b,c} \epsilon_{abc}\;
\left( {\partial({\ddot \psi}^{(2)}_{,a},
\psi^{(1)}_{,b},X_c) \over \partial (X_1,X_2,X_3)} +
{\partial({\psi}^{(2)}_{,a}, {\ddot
\psi}^{(1)}_{,b},X_c) \over \partial (X_1,X_2,X_3)}\right)
$$
$$
+ \sum_{a,b,c} {1 \over 2} \epsilon_{abc}\;{\partial({\ddot
\psi}^{(1)}_{,a},\psi^{(1)}_{,b},\psi^{(1)}_{,c}) \over \partial
(X_1,X_2,X_3)}
- \Lambda \; III(\psi^{(1)}_{,i,k}) \Bigr\rbrace = 0  \;\;.
\eqno(4d)
$$
Each order (starting from
$\varepsilon^1$) contains a term where a linear
operator acts on the different potentials $\psi^{(1)}$, $\psi^{(2)}$,
$\psi^{(3)}$. Similarily, a quadratic and a cubic operator
acts on quadratic or cubic invariants formed by
second derivatives of
$\psi^{(1)}$. However, at the third order, there appears
a quadratic term describing the interaction of linear and
quadratic perturbations. Higher orders will subsequently add such
interaction terms between the potentials at different or equal
orders.

We now derive the solutions of the equations (4)
in the case of a ``flat'' background. I present a simplified
derivation for a special class of initial conditions which is
relevant for the modeling of large-scale structure. Therefore,
it is illustrative to derive all orders with the same
restriction and procedure. The reader who
is interested in applications of the solutions only may proceed
directly to section {\bf 4}.
}

\vfill\eject

\noindent{\rmm 3. Solution of the perturbation equations}
\smallskip
{\rmm on an Einstein-de Sitter background}
\bigskip\bigskip
{\rmb
\noindent
In what follows we make use of a restriction of the
initial conditions to simplify the derivations.
We require that, initially, the peculiar-velocity $\vec u(\vec
X,t_0)$ be proportional to the peculiar-acceleration
$\vec w(\vec X,t_0)$:
$$
\vec u (\vec X,t_0) = \vec w (\vec X,t_0) t_0 \;\;\;,\eqno(5)
$$
where we
have defined the fields as usual (compare Peebles 1980, B92).
This restriction has proved to be appropriate for the purpose of
modeling large-scale structure since, for irrotational
flows, the peculiar-velocity field tends to be parallel to the
gravitational peculiar-field strength after some time.
The reason for this tendency is related to the existence of growing
and decaying perturbations, the growing part supports the
tendency to parallelity.
The assumption of irrotationality should be adequate down to the
non-linearity scale. However, besides the possibility of
non-linearly enhanced primordial vorticity (B92), shell-crossings
generate vorticity on scales below the non-linearity scale (see, e.g.,
Doroshkevich 1973, Chernin 1993 and ref. therein).
We should keep this in mind, since a third order approximation
actually should give a proper description of these smaller scales.
The treatment of vorticity is no longer academic, but might play a
decisive role for the dynamics within clusters of galaxies.
Also, decaying solutions should be considered with some care in the
non-linear regime. They couple to the growing solution, the role of
decaying and growing solutions in an expanding environment can be
interchanged in a collapsing environment as was discussed by
Gurevich and Zybin (1988b).
We note that the restriction (5)
is commonly used in the literature.
The tendency of the flow expressed
by (5) has been
proved for general first order (Bildhauer \& Buchert 1991) and a large
class of
second order irrotational flows (BE93). Note that in the case (5) we
have
to give one initial potential only, whereas the general initial value
setting for irrotational flows would require two. Any restriction
of this type simplifies the calculations enormously, but it implies the
restriction to irrotational flows.
(An alternative restriction of this type has been discussed in B92 and
BE93).

\medskip
We shall use in the following the initial peculiar-velocity potential
${\cal S}$ defined as $\vec u (\vec X,t_0) =: \nabla_0 {\cal S}(\vec
X)$. The initial peculiar-gravitational potential $\phi$,
$\vec w (\vec X,t_0) =:-\nabla_0 \phi (\vec X)$ is related to it as
${\cal S} = - \phi t_0$ (eq. (5)).
Consequently, we
seek solutions of the equations (4) of the form:
$$
\eqalignno{
\psi^{(1)} &= q_z (t) \;{\cal S}^{(1)} (\vec X) \;\;\;,&(6a)\cr
\psi^{(2)} &= q_{zz} (t)\; {\cal S}^{(2)} (\vec X) \;\;\;,&(6b)\cr
\psi^{(3)} &= q_{zzz} (t)\; {\cal S}^{(3)} (\vec X) \;\;\;,&(6c)\cr}
$$
where the potentials ${\cal S}^{(1)}, {\cal S}^{(2)}$ and ${\cal
S}^{(3)}$ have to be related to the initial condition $\cal S$.
We shall see that this relation will require the solution of
elliptic boundary value problems expressing non-locality of
the solutions.
Formally we require $q_z (t_0) =0$, $q_{zz} (t_0) =0$, $q_{zzz} (t_0)
=0$.
\bigskip\medskip
\noindent
{\rma 3.1. The zero order solution}
\medskip
\noindent
One class of
solutions of the Euler-Poisson system (2) is formed by
the homogeneous and
isotropic Friedman-Lema\^\i tre cosmologies:
$$
{\vec f}_H (\vec X,t) = a(t) \vec X \;. \eqno(7a)
$$
Inserting ${\vec f}_H$ into the eqations (2), we obtain
for the function $a(t)$ the zero order equation (4a).
Its general solution is given by solutions of
Friedman's differential
equation as an integral of (4a):
$$
{{\dot a}^2 + const \over a^2} = {8 \pi G \rho_H + \Lambda \over 3} \; ,
\eqno(7b)
$$
where $\rho_H = \ueber{\rho}{o}_H a^{-3}$ is the background density.
Henceforth we restrict all considerations to the Einstein-de Sitter
case ($const=0, \Lambda=0$). Then, the zero order solution reads:
$$
a(t) = \left( t \over t_0 \right)^{2 \over 3} \;\;\;. \eqno(7c)
$$
For convenience, we
shall express all time-dependent coefficients in terms of the
solution $a$.
The constant $-4 \pi G \ueber{\rho}{o}_H$ will be written in terms of
its value $-{2 \over 3 t_0^2}$ for the solution (7c).

\bigskip\medskip
\noindent
{\rma 3.2. The first order solution}
\medskip
\noindent
By virtue of the ansatz (6a),
equation (4b) simplifies to the following
equation (in the sequel we put $\Lambda=0$ and use (3c) and (7c)):
$$
\left({\ddot q}_z + 2 {\ddot a \over a} q_z \right) \Delta_0 {\cal
S}^{(1)} = {1 \over a^2} (- {2 \over 3 t_0^2})\;
\ueber{\delta}{0} \;\;.\eqno(8)
$$
Using Poisson's equation for the initial potential $\phi$ and the
relation ${\cal S}= - \phi t_0$ we have:
$$
\Delta_0 {\cal S} = - {2 \over 3 t_0} \ueber{\delta}{o}
\;\;.\eqno(8a)
$$
Hence, solutions of (8) can be found as solutions
of the linear ordinary differential equation:
$$
{\ddot q}_z + 2 {\ddot a \over a} q_z = {1 \over a^2 t_0^2} =: {\cal
G}^{(1)}(t) \eqno(8b)
$$
with:
$$
\Delta_0 {\cal S}^{(1)} = \Delta_0 {\cal S} \; t_0 =
I({\cal S}_{,i,k}) \;t_0 \;\;\;. \eqno(8c)
$$
The solution to (8b) consists of two linearly independent
solutions of the homogeneous part:
$$
q_1^{(1)} = C_1^{(1)} a^2 \;\;,\;\; q_2^{(1)} = C_2^{(1)} a^{-{1 \over
2}} \;\;, \eqno(9a)
$$
and a particular solution:
$$
q_p^{(1)} = C^{(1)} \left(q_1^{(1)} \int^t q_2^{(1)} {\cal
G}^{(1)} dt
- q_2^{(1)} \int^t q_1^{(1)} {\cal G}^{(1)} dt \right) = C^{(1)} a
\;\;\;.\eqno(9b)
$$
The coefficient $C^{(1)}$ is found by inserting $q_p^{(1)}$ into (8b);
the coefficients $C_1^{(1)}$ and $C_2^{(1)}$ are found by the
requirement that the coefficient functions of the peculiar-velocity
and -acceleration equal to $1$ at $t=t_0$ for the solution
$q^{(1)} = q_1^{(1)}+q_2^{(1)}+q_p^{(1)}$.
Restricting the general solution $q^{(1)}$
according to (5) we find (B92):
$$
q_z = {3 \over 2} (a^2 - a) \;\;\;. \eqno(10)
$$
A discussion of the uniqueness of this solution related to the
use of ${\cal S}t_0$ instead of ${\cal S}^{(1)}$ is to be found in
APPENDIX B.

\bigskip\medskip
\noindent
{\rma 3.3. The second order solution}
\medskip
\noindent
Inserting the ansatz (6b) into the equation (4c) we find:
$$
\left( {\ddot q}_{zz} + 2 {\ddot a \over a} q_{zz} \right) \Delta_0{\cal
S}^{(2)} = - \left({{\ddot q}_z q_z \over a} + {q_z^2 \over 2}
{\ddot a \over a^2} \right) 2 II({\cal S}^{(1)}_{,i,k})\;\;.\eqno(11)
$$
Solutions to (11) can be found by solving the linear ordinary
differential equation (inserting the first order solution $q_z$):
$$
{\ddot q}_{zz} + 2 {\ddot a \over a} q_{zz} = - \left(
{{\ddot q}_z q_z \over a} + {q_z^2 \over 2} {\ddot a \over a^2}
\right) =: {\cal G}^{(2)} (t)\;\;, \eqno(11a)
$$
with:
$$
\Delta_0 {\cal S}^{(2)} = 2 II({\cal S}^{(1)}_{,i,k})
\;\;.\eqno(11b)
$$
The solution to (11a) consists of two linearly independent
solutions of the homogeneous part:
$$
q_1^{(2)} = C_1^{(2)} a^2 \;\;,\;\; q_2^{(2)} = C_2^{(2)} a^{-{1 \over
2}} \;\;,\eqno(12a)
$$
and a particular solution:
$$
q_p^{(2)} = C^{(2)} \left(q_1^{(2)} \int^t q_2^{(2)} {\cal
G}^{(2)} dt
- q_2^{(2)} \int^t q_1^{(2)} {\cal G}^{(2)} dt \right)
= C^{(2)} (-{3 \over 4} + {3 \over 4} a^{-2})
\;\;\;.\eqno(12b)
$$
The coefficient $C^{(2)}$ is found by inserting $q_p^{(2)}$ into (11a);
the coefficients $C_1^{(2)}$ and $C_2^{(2)}$ are found by the
requirement that the coefficient functions of the peculiar-velocity
and -acceleration vanish at $t=t_0$ for the solution
$q^{(2)} = q_1^{(2)}+q_2^{(2)}+q_p^{(2)}$.
We find for the restriction (5) the following result (compare BE93,
section {\bf 5}):
$$
q_{zz} = \left({3 \over 2}\right)^2 (-{3 \over 14} a^3 + {3 \over 5} a^2
- {1 \over 2} a + {4 \over 35} a^{-{1 \over 2}}) \;\;\;. \eqno(13)
$$

\bigskip\medskip
\noindent
{\rma 3.4. The third order solution - longitudinal part}
\medskip
\noindent
Inserting the ansatz (6c) into the equation (4d) we find:
$$
\eqalignno{
\left( {\ddot q}_{zzz} + 2 {\ddot a \over a} q_{zzz} \right)
\Delta_0 {\cal
S}^{(3)} = &- \left({\ddot a \over a^2} q_{zz} q_z + {1 \over a}
( {\ddot q}_{zz} q_z + {\ddot q}_z q_{zz} )\right)
\sum_{a,b,c} \epsilon_{abc}\;
{\partial({\cal S}^{(2)}_{,a},{\cal S}^{(1)}_{,b},X_c)
\over \partial (X_1,X_2,X_3)} \cr
&- \left( {1 \over a^2} {\ddot q}_z q_z^2 \right) \; 3 III({\cal
S}^{(1)}_{,i,k}) \;\;\;. &(14) \cr}
$$
To obtain solutions of (14) we use the linearity of Poisson's
equation and split the potential ${\cal S}^{(3)}$ into a part
${\cal S}^{(3a)}$ generating the cubic source term, and a part
${\cal S}^{(3b)}$ generating the quadratic source term of the
interaction of first and second order perturbations.
We then have to solve separately the linear ordinary
differential equation (inserting the first order solution $q_z$):
$$
{\ddot q}_{zzz} + 2 {\ddot a \over a} q_{zzz} = -
{1 \over a^2} {\ddot q}_z q_z^2 =: {\cal G}^{(3a)} (t)\;\;, \eqno(14a)
$$
with:
$$
\Delta_0 {\cal S}^{(3a)} =
3 III({\cal S}^{(1)}_{,i,k}) \;\;\;, \eqno(14b)
$$
and the linear ordinary differential equation (inserting the first and
second order solutions $q_z$ and $q_{zz})$:
$$
{\ddot q}_{zzz} + 2 {\ddot a \over a} q_{zzz} = -
{\ddot a \over a^2} q_{zz} q_z + {1 \over a} ( {\ddot q}_{zz} q_z
+ {\ddot q}_z q_{zz} ) =: {\cal G}^{(3b)} (t)\;\;, \eqno(14c)
$$
with:
$$
\Delta_0 {\cal S}^{(3b)} =
\sum_{a,b,c} \epsilon_{abc}\;
{\partial({\cal S}^{(2)}_a,{\cal S}^{(1)}_b,X_c)
\over \partial (X_1,X_2,X_3)}
\;\;\;.\eqno(14d)
$$
A general solution to (14a) or (14c), respectively,
consists of two linearly independent
solutions of the homogeneous part:
$$
q_1^{(3a,b)} =
C_1^{(3a,b)} a^2 \;\;,\;\; q_2^{(3a,b)} = C_2^{(3a,b)} a^{-{1 \over
3}} \;\;,\eqno(15a,b)
$$
and a particular solution for each source term:
$$
q_p^{(3a)} = C^{(3a)} \left(q_1^{(3a)} \int^t q_2^{(3a)} {\cal
G}^{(3a)} dt
- q_2^{(3a)} \int^t q_1^{(3a)} {\cal G}^{(3a)} dt
\right)
$$
$$
= C^{(3a)} {9 \over 4}
(- {2 \over 3} a + 1 - {1 \over 3} a^{-2})
\;\;\;.\eqno(16a)
$$
$$
q_p^{(3b)} = C^{(3b)} \left(q_1^{(3b)} \int^t q_2^{(3b)} {\cal
G}^{(3b)} dt
- q_2^{(3b)} \int^t q_1^{(3b)} {\cal G}^{(3b)} dt
\right)
$$
$$
= C^{(3b)} {9 \over 4}({5 \over 7} a - {11 \over 10} + {1 \over
2} a^{-2} - {4 \over 35} a^{-{5 \over 2}})
\;\;\;.\eqno(16b)
$$
The coefficients $C^{(3a,b)}$ are
found by inserting $q_p^{(3a,b)}$ into (16a) or (16b), respectively;
the coefficients $C_1^{(3a,b)}$ and $C_2^{(3a,b)}$ are found as in
the second order case by the
requirement that the coefficients of the initial peculiar-velocity and
-acceleration vanish for the general solutions.
I find for the restriction (5) the following result:
$$
q_{zzz}^{a} = \left({3 \over 2}\right)^3
(-{1 \over 9} a^4 + {3 \over 7} a^3 - {3 \over 5} a^2 +
{1 \over 3} a - {16 \over 315} a^{-{1 \over 2}}) \;\;\;; \eqno(17a)
$$
$$
q_{zzz}^{b} = \left({3 \over 2}\right)^3
({5 \over 42} a^4 - {33 \over 70} a^3 + {7 \over 10} a^2
- {1 \over 2} a + {4 \over 35} a^{1 \over 2}
+ {4 \over 105} a^{-{1 \over 2}}) \;\;\;. \eqno(17b)
$$

\vfill\eject
\noindent
{\rma 3.5. The third order solution - transverse part}
\medskip
\noindent
Inserting the longitudinal ansatz
$$
{\vec f}^L = \nabla_0 \left( a {\vert\vec X\vert^2 \over 2} +
q_z {\cal S}^{(1)} + q_{zz} {\cal S}^{(2)} + q_{zzz}^{a} {\cal S}^{(3a)}
+ q_{zzz}^{b} {\cal S}^{(3b)} \right)       \eqno(18)
$$
into the equations (2a,b,c), we find restrictions
on the initial conditions. These restrictions are derived in
APPENDIX A. Note that, instead of the equations (2a,b,c) for the
irrotationality of the gravitational field strength, we can use
the corresponding equations for the irrotationality of the velocity
in the case of irrotational flows (see the LEMMA proved in BE93), which
we shall do.

\noindent
The restrictions only arise at the third order level. This implies that
the third order solution is not purely longitudinal, it is not
fully covered by the ansatz (18). In the following we seek
a transverse part of the third order solution by extending the
ansatz (18):
$$
\vec f = {\vec f}^L + {\vec f}^T \;\;;\;\;{\vec f}^T := q^c_{zzz}
\vec \Xi \;\;;\;\;\vec \Xi := -
\nabla_0 \times {\vec {\cal S}}^{(3c)} \;\;.\eqno(19a)
$$
We have introduced the vector potential ${\vec {\cal S}}^{(3c)}$,
on which we impose the following gauge condition (compare APPENDIX B):
$$
\nabla_0 \cdot {\vec {\cal S}}^{(3c)} \;\;\;. \eqno(19b)
$$
With the ansatz (19a)
we generate no additional equations to be fulfilled
in the source equation (2d). The integrability conditions for the
velocity (BE93: equations (5d,e,f)) only yield an equation of the order
$\varepsilon^3$ as expected ($i,j,k=1,2,3$ cyclic ordering):
$$
\varepsilon^3 \Bigl\lbrace a \left( \dot a q^c_{zzz} - a {\dot
q}^c_{zzz} \right)
\left(\nabla_0 \times (-\nabla_0 \times {\vec {\cal S}}^{(3c)})\right)_k
$$
$$
+ a \left( {\dot q}_{zz} q_z - {\dot q}_z q_{zz} \right)
\epsilon_{pq \lbrack j} {\partial ({\cal S}^{(2)}_{,i \rbrack},
{\cal S}^{(1)}_{,p},X_q) \over \partial(X_1,X_2,X_3)} \Bigr\rbrace \;
= 0 \;\;\;.\eqno(20a,b,c)
$$
Using the vector identity $\nabla_0 \times (\nabla_0 \times {\vec
{\cal S}}^{(3c)}) = \nabla_0 (\nabla_0 \cdot {\vec {\cal S}}^{(3c)}) -
\Delta_0 {\vec {\cal S}}^{(3c)}$ and (19b), we find solutions
of the equations (20) by solving the linear ordinary
differential equation (inserting the first and second order solutions
$q_z$ and $q_{zz}$):
$$
{\dot q}^c_{zzz} - {\dot a \over a} q^c_{zzz} = -
{1 \over a} \left( {\dot q}_{zz} q_z - {\dot q}_z q_{zz} \right)
=: {\cal G}^{(3c)} (t)\;\;, \eqno(21a)
$$
with:
$$
(\Delta_0 {\vec {\cal S}}^{(3c)})_k =
\epsilon_{pq \lbrack j} {\partial ({\cal S}^{(2)}_{,i \rbrack},
{\cal S}^{(1)}_{,p},X_q) \over \partial(X_1,X_2,X_3)}
\;\;\;. \eqno(21b,c,d)
$$
A general solution of (21a) is given by:
$$
q^c_{zzz} = {1 \over M(t)} \left( \int_{t_0}^t {\cal G}^{(3c)}(t) M(t)
dt \; + C^{(3c)} \right)
\;\;, \eqno(22)
$$
with the integrating factor $M(t) = e^{-\int_{t_0}^t {\dot a \over a}
dt}$.
The coefficient $C^{(3c)}$ is
found by the requirement $q^c_{zzz} (t_0) = 0$.
We finally obtain:
$$
q_{zzz}^{c} = \left({3 \over 2}\right)^3
({1 \over 14} a^4 - {3 \over 14} a^3 + {1 \over 10} a^2 +
{1 \over 2} a - {4 \over 7} a^{{1 \over 2}} +
{4 \over 35} a^{-{1 \over 2}}) \;\;\;. \eqno(23)
$$

}

\vfill\eject

\noindent{\rmm 4. Result and Discussion}
\bigskip\smallskip
{\rmb
\noindent
{\rma 4.1. The solution}
\bigskip
{\bf THEOREM}
\medskip
\noindent
With a superposition ansatz for Lagrangian perturbations of an
Einstein-de Sitter background we have obtained
the following family of trajectories $\vec x = \vec f
(\vec X,a)$ as {\bf irrotational}
solution of the Euler-Poisson system up to the third
order in the perturbations from homogeneity. The general set of
initial conditions $(\phi (\vec X), {\cal S} (\vec X))$ is restricted
according to ${\cal S} = - \phi t_0$.
(The parameter $\varepsilon$ is considered as the
amplitude of the initial fluctuation field; $a(t) = (t/t_0)^{2/3}$,
$i,j,k=1,2,3$ with cyclic ordering):
$$
\vec f = a \; \vec X \;+\;
q_z (a) \; \nabla_0 {\cal S}^{(1)} (\vec X) \;+\; q_{zz}
(a) \; \nabla_0 {\cal S}^{(2)} (\vec X)
$$
$$
+\; q_{zzz}^{a} (a) \;
\nabla_0 {\cal S}^{(3a)} (\vec X) \;+\; q_{zzz}^{b} (a) \;
\nabla_0 {\cal S}^{(3b)} (\vec X) \;-\;
q_{zzz}^{c} (a) \; \nabla_0 \times {\vec {\cal S}}^{(3c)} (\vec X)
\;\;,\eqno(24)
$$
with:
$$
\eqalignno{
q_z &= \left({3 \over 2}\right) (a^2 - a) \;\;\;, &(24a) \cr
q_{zz} &= \left({3 \over 2}\right)^2
(-{3 \over 14} a^3 + {3 \over 5} a^2 - {1 \over 2}
a + {4 \over 35} a^{-{1 \over 2}}) \;\;\;, &(24b) \cr
q_{zzz}^{a} &= \left({3 \over 2}\right)^3
(-{1 \over 9} a^4 + {3 \over 7} a^3 - {3 \over 5} a^2
+ {1 \over 3} a - {16 \over 315} a^{-{1 \over 2}}) \;\;\;,&(24c) \cr
q_{zzz}^{b} &= \left({3 \over 2}\right)^3
({5 \over 42} a^4 - {33 \over 70} a^3 + {7 \over 10} a^2
- {1 \over 2} a + {4 \over 35} a^{1 \over 2}
+ {4 \over 105} a^{-{1 \over 2}}) \;\;\;, &(24d) \cr
q_{zzz}^{c} &= \left({3 \over 2}\right)^3
({1 \over 14} a^4 - {3 \over 14} a^3 + {1 \over 10} a^2
+ {1 \over 2} a - {4 \over 7} a^{1 \over 2}
+ {4 \over 35} a^{-{1 \over 2}}) \;\;\;, &(24e) \cr}
$$
and:
$$
\eqalignno{
&\Delta_0 {\cal S}^{(1)} = I({\cal S}_{,i,k}) \;t_0 \;\;\;,&(24f) \cr
&\Delta_0 {\cal S}^{(2)} = 2 II({\cal S}^{(1)}_{,i,k})
\;\;\;, &(24g) \cr
&\Delta_0 {\cal S}^{(3a)} =
3 III({\cal S}^{(1)}_{,i,k}) \;\;\;, &(24h) \cr
&\Delta_0 {\cal S}^{(3b)} =
\sum_{a,b,c} \epsilon_{abc}\;
{\partial({\cal S}^{(2)}_{,a},{\cal S}^{(1)}_{,b},X_c)
\over \partial (X_1,X_2,X_3)} \;\;\;, &(24i) \cr
(&\Delta_0 {\vec {\cal S}}^{(3c)})_k =
\epsilon_{pq \lbrack j} {\partial ({\cal S}^{(2)}_{,i \rbrack},
{\cal S}^{(1)}_{,p},X_q) \over \partial(X_1,X_2,X_3)}
\;\;\;. &(24j,k,l) \cr}
$$
\vfill\eject
\noindent
{\sl REMARKS:}
\smallskip\noindent
The potential ${\cal S}^{(3b)}$ and the vector potential
${\vec {\cal S}}^{(3c)}$
generate interaction among the
first and second order perturbations. The general interaction
term is not purely
longitudinal.
In order to satisfy the Euler-Poisson system with the
longitudinal part only, we have to respect the following constraint
(APPENDIX A):
$$
\nabla_0 {\cal S}^{(2)} =
{\cal W} (\nabla_0 {\cal S}^{(1)}) \;\;\;. \eqno(24m)
$$
The potential ${\cal S}^{(3b)}$ generates the symmetric part, whereas
the vector potential ${\vec {\cal S}}^{(3c)}$ generates the
anti-symmetric part of the interaction.
\bigskip\noindent
{\bf Proof:} The proof follows by inserting the solution (24) into
the perturbation equations (4), which has been done using the
algebraic manipulation system {\sl REDUCE}.
\bigskip\noindent
We now discuss some issues which are related to the practical use
of the solution (24) as a model for the evolution of large-scale
structure.

\bigskip\medskip
\noindent
{\rma 4.2. The construction of local forms}
\medskip
\noindent
The usefulness of {\it local} approximations has been pointed out
by Nusser \etal (1991) who apply the ``Zel'dovich approximation''
as a tool for locally reconstructing the density field from
observed peculiar-velocity data. Contrary, solutions of Newtonian
equations for the evolution of self-gravitating ``dust'' continua
are {\it non-local}, since, e.g., the gravitational potential has to
be a solution of the (elliptic) Poisson equation. This solution
involves integrals over large space regions. As far as the first
order Lagrangian perturbation solution is concerned, this is not a
contradiction: without loss of generality, i.e., without change of
physical quantities like the density contrast or the divergence of the
peculiar-velocity, we can use the initial condition ${\cal S}t_0$
instead of
${\cal S}^{(1)}$ (APPENDIX B).
With this reduction the iterative procedure of
constructing
the displacement vectors in (24) simplifies: the source
terms are (except the interaction terms (24i-l)) completely expressible
in terms of second derivatives of the initial potential $\cal S$.

The situation
is more delicate at higher order levels. We no longer are able
to write second or higher order terms as a local approximation.
Although it is computationally simple to obtain the
displacement vectors
in (24) by employing Poisson solvers (see {\bf 4.3}), it is useful to
know which
classes of initial conditions admit the construction of local forms.
With such closed form expressions the study of special solutions is
explicitly feasible without using a numerical Poisson solver.

\medskip
According to {\sl COROLLARY 1} proved in (BE93), a local form can
be obtained for second order displacements. It reads:
$$
\nabla_0 {\cal S}^{(2)}  = \nabla_0
{\cal S} \left( \Delta_0 {\cal S} \right)
- \left( \nabla_0 {\cal S} \cdot \nabla_0 \right) \nabla_0 {\cal S}
\;\;;\;\; \nabla_0 {\cal S} \times
\Delta_0 \nabla_0 {\cal S} = \vec 0 \;\,.
\eqno(25a,b,c,d)
$$
The local form (25a) is constructed such that its divergence
agrees with the source term in (24g), its curl is, however, in
general non-zero, it only vanishes if (25b-d) is statisfied.
The latter equations express the fact that the form (25a) cannot
be used in general, constraints have to be obeyed.

\noindent
Similarily, we can ask for a local vector form whose divergence
agrees with the third principal
scalar invariant of $({\cal S}_{,i,k})$, which is
the source term in equation (24h). Indeed, the following expression
has the required property:
\bigskip
{\bf COROLLARY 1}
\medskip\noindent
The vector $\nabla_0 {\cal S}^{(3a)}$ with the components
$$
(\nabla_0 {\cal S}^{(3a)})_{k}  =  \sum_i (\nabla_0 {\cal S}^{(1)})_{,i}
J^S_{i,k}
\eqno(26a)
$$
has the property:
$$
\Delta_0 {\cal S}^{(3a)} =
3 III({\cal S}^{(1)}_{,i,k}) \;\;\;,
$$
where $J^S_{i,k}$ are the subdeterminants of the tensor
$({\cal S}^{(1)}_{,i,k})$.
The following constraints have to be satisfied in order that
$\nabla_0 {\cal S}^{(3a)}$ is curl-free:
$$
\sum_i (\nabla_0 {\cal S}^{(1)})_{,i} J^S_{i,\lbrack k,j\rbrack} = 0
\;\;\;,\;\;\;k \ne j \;\;\;. \eqno(26b,c,d)
$$
(The proof is done by explicit verification).

\bigskip\noindent
The source term in (24i) which describes the longitudinal
part of the interaction
of first and second order perturbations has a similar structure
as the second order source term. We can construct a local form by
analogy. The integral of the interaction source term in equation (24i)
reads (BE93, COROLLARY 1):
$$
\nabla_0 {\cal S}^{(3b)} =
$$
$$
\lambda_1 \left( \nabla_0 {\cal S}^{(2)} ( \Delta_0 {\cal S}^{(1)} )
- ( \nabla_0 {\cal S}^{(2)} \cdot \nabla_0 ) \nabla_0 {\cal
S}^{(1)} \right) + \lambda_2 \left(
\nabla_0 {\cal S}^{(1)} ( \Delta_0 {\cal S}^{(2)}
- \nabla_0 {\cal S}^{(1)} \cdot \nabla_0 ) \nabla_0 {\cal
S}^{(2)} \right) \;\;. \eqno(27a)
$$
We have taken the linear combination
of the two possible integrals as a
general integral, where we have to assure $\lambda_1 + \lambda_2 = 1$.
In order to satisfy the requirement that the vector (27a) be a solution
of the Poisson equation (24i), we have to assure that it is curl-free
which implies (BE93, COROLLARY 1):
$$
\lambda_1 \left(
\nabla_0 {\cal S}^{(2)} \times \Delta_0 \nabla_0 {\cal S}^{(1)} \right)
\;+\; \lambda_2 \left(
\nabla_0 {\cal S}^{(1)} \times \Delta_0 \nabla_0 {\cal S}^{(2)} \right)
\;=\; \vec 0 \;\;. \eqno(27b,c,d)
$$
\bigskip\noindent
We finally give an integral expression for the source terms in
(24j,k,l) which describe the transverse part of the interaction
of first and second order perturbations:

\bigskip
{\bf COROLLARY 2}
\medskip\noindent
The vector $\vec \Xi$:
$$
\vec \Xi : = - \nabla_0 \times {\vec {\cal S}}^{(3c)} =
$$
$$
\mu_1 \left( (\nabla_0 {\cal S}^{(2)} \cdot \nabla_0 )
\nabla_0 {\cal S}^{(1)} \right) +
\mu_2 \left( -(\nabla_0 {\cal S}^{(1)} \cdot \nabla_0 )
\nabla_0 {\cal S}^{(2)} \right)   \eqno(28a,b,c)
$$
has the property:
$$
(\nabla_0 \times \vec \Xi)_i =
\epsilon_{pq \lbrack j} {\partial ({\cal S}^{(2)}_{,i \rbrack},
{\cal S}^{(1)}_{,p},X_q) \over \partial(X_1,X_2,X_3)} \;\;,\; i \ne j
\;\;\;.
$$
We have taken the linear combination
of the two possible integrals as a
general integral, where we have to assure $\mu_1 + \mu_2 = 1$.
In order to satisfy the requirement that the vector components (28a,b,c)
be solutions of the Poisson equations (24j,k,l), we have to assure that
the displacement vector $\vec \Xi$ can be represented in terms of the
vector potential
${\vec {\cal S}}^{(3c)}$: $\vec \Xi = -\nabla_0 \times
{\vec {\cal S}}^{(3c)}$, i.e., the vector field $\vec \Xi$ has to be
source-free:
$$
\mu_1 \nabla_0 \cdot
\left( (\nabla_0 {\cal S}^{(2)} \cdot \nabla_0 )
\nabla_0 {\cal S}^{(1)} \right) -
\mu_2 \nabla_0 \cdot \left( (\nabla_0 {\cal S}^{(1)} \cdot \nabla_0 )
\nabla_0 {\cal S}^{(2)} \right) = 0  \;\;.\eqno(28d)
$$
(The proof is done by explicit verification).

\bigskip\noindent
In special cases it is possible to construct the four potentials ${\cal
S}^{(3b)}$ and $({\vec {\cal S}}^{(3c)})_i$
by fixing the parameters $\lambda_1$, $\lambda_2$, $\mu_1$, and $\mu_2$
suitably,
thus fulfilling the constraints (27b) and (28d,e,f); in some cases
the integration freedom (the curl of some vector in (27a), and the
gradient of some scalar in (28a,b,c)) has to be used in order to fulfil
the constraints, (see Buchert \etal 1993c).
\vfill\eject
\noindent
{\rma 4.3. Practical realization}
\smallskip
{\rma $\;\;\;$and comparison with numerical simulations}
\bigskip\smallskip
\noindent
In order to realize the presented solution for {\it generic} initial
conditions the following procedure is appropriate.

\noindent
Let us start with a Gaussian random
density contrast
field $\ueber{\delta}{o} (\vec X)$ as initial condition.
It is most convenient
to express this field in terms of its discrete Fourier transform
(see, e.g., Bertschinger 1992).
Henceforth, we denote
its Fourier components by $\hat\delta$, the Fourier sums may extend over
$n$ wave vectors $\vec K = (K_1, K_2, K_3)$.
 From the density perturbation we construct the initial
peculiar-velocity potential in $K-$space
using FFT (``Fast Fourier Transform''),
here written for the ``flat'' background:
$$
\hat{\cal S} (\vec K) = - {2 \over 3 t_0} \Delta_0^{-1}(\vec K)
\hat\delta (\vec K) \;\;.\eqno(29)
$$
 From $\hat{\cal S}(\vec K)$ we are able to construct all displacement
vectors which are
generated by second derivatives of ${\cal S}$ with respect to
Lagrangian coordinates. This can be done by
simply multiplying and summing
the components $K_1, K_2, K_3$ of all $n$ wave vectors ${\vec
K}$ in $K-$space.
We first use the following property of Fourier transforms:
$$
\widehat{\nabla_0 {\cal S} (\vec X)} (\vec K) =
i \vec K \; \hat {\cal S} (\vec K) \;\;; \eqno(30a)
$$
$$
\widehat {({\cal S}_{,i,k}) (\vec X)} (\vec K) =
- K_{i}K_{k} \; \hat {\cal S} (\vec K) \;\;.\eqno(30b)
$$
Transforming the gradient
back to the $X-$space we obtain the displacement vector of the
first order perturbation. Transforming all $6$ different Fourier
components of the
symmetric tensor gradient $\widehat{({\cal S}_{,i,k})}$
back to the $X-$space we
are able to construct the
three principal scalar invariants of it:
$$
\eqalignno{
I({\cal S}_{,i,k})
&= {\cal S}_{,1,1} + {\cal S}_{,2,2} + {\cal S}_{,3,3} \;\;,
&(31a) \cr
II({\cal S}_{,i,k})
&= {\cal S}_{,1,1}{\cal S}_{,2,2} + {\cal S}_{,1,1}{\cal S}_{,3,3}
+ {\cal S}_{,2,2}{\cal S}_{,3,3} - {\cal S}_{,1,2}{\cal S}_{,2,1}
- {\cal S}_{,1,3}{\cal S}_{,3,1} - {\cal S}_{,2,3}{\cal S}_{,3,2}
\;\; &(31b) \cr
III({\cal S}_{,i,k})
&= {\cal S}_{,3,3}{\cal S}_{,2,2}{\cal S}_{,1,1} -
   {\cal S}_{,3,3}{\cal S}_{,2,1}{\cal S}_{,1,2} -
   {\cal S}_{,3,2}{\cal S}_{,2,3}{\cal S}_{,1,1} + \cr
&\;\;\;\;\;{\cal S}_{,3,2}{\cal S}_{,2,1}{\cal S}_{,1,3} +
           {\cal S}_{,3,1}{\cal S}_{,2,3}{\cal S}_{,1,2} -
           {\cal S}_{,3,1}{\cal S}_{,2,2}{\cal S}_{,1,3} \;\;.
&(31c) \cr}
$$
 From the invariants (31b) and (31c) we construct the potentials
${\cal S}^{(2)}$ and ${\cal S}^{(3a)}$ by using again FFT in analogy
to (29).
Note, that for the displacement vectors in (24) we need the
gradients of these potentials which can be constructed again by
multiplying with $i \vec K$ in $K-$space.

\noindent
To construct the interaction terms, we have to use the components of
$\nabla_0 {\cal S}^{(2)}$ and mix them up with the components of
$\nabla_0 {\cal S}^{(1)}$. This is, of course, only possible after
the construction of ${\cal S}^{(2)}$:
After building the mixed invariants, for the longitudinal interaction
term:
$$
II^L_{mix}
= {\cal S}^{(2)}_{,1,1}{\cal S}_{,2,2}
+ {\cal S}^{(2)}_{,1,1}{\cal S}_{,3,3}
+ {\cal S}^{(2)}_{,2,2}{\cal S}_{,3,3}
- {\cal S}^{(2)}_{,1,2}{\cal S}_{,2,1}
- {\cal S}^{(2)}_{,1,3}{\cal S}_{,3,1}
- {\cal S}^{(2)}_{,2,3}{\cal S}_{,3,2}
$$
$$
+ {\cal S}_{,1,1}{\cal S}_{,2,2}^{(2)}
+ {\cal S}_{,1,1}{\cal S}_{,3,3}^{(2)}
+ {\cal S}_{,2,2}{\cal S}_{,3,3}^{(2)}
- {\cal S}_{,1,2}{\cal S}_{,2,1}^{(2)}
- {\cal S}_{,1,3}{\cal S}_{,3,1}^{(2)}
- {\cal S}_{,2,3}{\cal S}_{,3,2}^{(2)}
\;\;, \eqno(31d)
$$
and for the transverse interaction terms:
$$
(II^T_{mix})_i
= {\cal S}^{(2)}_{,i,j}{\cal S}_{,i,k}
+ {\cal S}^{(2)}_{,j,j}{\cal S}_{,j,k}
+ {\cal S}^{(2)}_{,k,j}{\cal S}_{,k,k}
- {\cal S}_{,i,j}{\cal S}_{,i,k}^{(2)}
- {\cal S}_{,j,j}{\cal S}_{,j,k}^{(2)}
- {\cal S}_{,k,j}{\cal S}_{,k,k}^{(2)} \;\;\;;
$$
$$
i,j,k = 1,2,3 \;\;\;cyclic \;\,ordering
\;\;, \eqno(31e,f,g)
$$
we are able to construct the four generating functions of the
interaction terms ${\cal S}^{(3b)}$ and ${\vec {\cal S}}^{(3c)}$ via
FFT, and from those functions the components of the displacement
vectors.
\bigskip
The presented model is currently being compared with numerical
simulations.
Two of the comparisons (Buchert \etal 1993a,b)
concern the cross-correlation of generic density
fields as predicted by the Lagrangian perturbation solutions with
N-body simulations. This is done
for various different power spectra similar to the work
by Coles \etal (1993).
Another comparison (Buchert \etal 1993c) performs a study of special
cluster models
using a tree-code. The purpose of the
latter work is to learn about the principal
effects of the different orders at high resolution.
In this study the local forms
discussed in {\bf 4.2} can be used, the models employed
are examples of third-order solutions which are
fully expressible in closed form without using a numerical
Poisson solver.

\bigskip\medskip
\noindent
{\rma 4.4. Discussion}
\medskip
\noindent
In the sequel all statements are understood in terms of the restriction
(5). Similar statements will hold for the general case.
\smallskip\noindent
The infinite sequence of a perturbation ansatz of the form:
$$
\vec f = a \vec X + \sum_{\ell=1}^{\infty} \varepsilon^{\ell} {\vec
p}^{(\ell)}
\eqno(32)
$$
contains in its spatial part integrals of
invariants which involve sums and products
of first derivatives of the perturbation fields $p^{\ell}_i$.
All these
invariants are at most cubic in the products. This fact implies that the
sequence (32) starts with those terms listed in (24) which grow
as $q_z$, $q_{zz}$ and $q_{zzz}^a$, the whole rest of the sequence
is made up of {\it interaction} terms resulting from products
of perturbations which are not directly expressible in terms of the
initial conditions.
In view of this it is natural to retain those three terms
as ``main body'' of the perturbation sequence for modeling the early
non-linear regime, thus truncating {\it
all} interaction terms. This way a truncated model
is obtained which is the simplest third order model
for the study of large-scale structure.
In the sequel we list the advantages of the truncated model
${\vec f}^{trunc} =
a \vec X + q_z \nabla_0 {\cal S}^{(1)} + q_{zz}
\nabla_0 {\cal S}^{(2)} + q_{zzz}^{a} \nabla_0 {\cal S}^{(3a)}$:
\bigskip

$\bullet$
The model respects
all three scalar invariants of the initial displacement tensor
$({\cal S}_{,i,k})$.
\smallskip
$\bullet$
This (and only this) model maps the intial condition
to the final stage directly without

$\;\;\;\;\;\;$the need to construct the
displacement vectors iteratively.
\smallskip
$\bullet$
The model is purely longitudinal.

\bigskip
However, the interaction terms (not contained in $\vec f^{trunc}$)
grow at the rates $q_{zzz}^b$ and $q_{zzz}^c$ comparable with
$q_{zzz}^a$.
The comparison with numerical simulations will show, whether
neglection of these terms is meaningful. Note, that a similar growth
rate does not imply similar importance for the clustering process,
because that depends on the detailed spatial structure of the source
terms in (24i-l) compared to the source term in (24h). The other
question,
whether the local forms discussed in section {\bf 4.2} might be
useful also
for generic initial conditions will also be resumed in future work.

\noindent
We finally emphasize that we can expect the Lagrangian
perturbation ansatz to apply until $a(t) \sigma (t_0) =
{\cal O}(1)$, where $\sigma (t_0) = {3 \over 2} \varepsilon$ is the
r.m.s. amplitude of the initial density contrast field. At this
``accumulation point''
all orders of the approximations result in displacements of the
same order as the first order displacements. Obviously, this
restriction is much less severe as the bound $\vert \delta \vert <
1$ in the Eulerian perturbation theory, since the density itself is not
bounded in the Lagrangian framework.

\vfill\eject
\noindent
{\bf Note:} A {\sl REDUCE} program of all equations in this paper
and the solution
can be obtained from the author via electronic mail:
{\sl TOB at ibma.ipp-garching.mpg.de}.

}

\bigskip\bigskip
\noindent
{\rma Acknowledgements:}
{\rmc
I would like to thank V.F. Mukhanov for interesting discussions
on gravitational multi-stream systems, S.F. Shandarin and A.L. Melott
for valuable
remarks and discussions. In particular, I am greatful to J. Ehlers
for clarifying comments on the manuscript and discussions.

\noindent
This work is supported by DFG (Deutsche Forschungsgemeinschaft).

}

\vfill\eject

\centerline{\rmm APPENDIX A}
\bigskip\bigskip
{\rmb
\noindent
In this APPENDIX the constraint equations resulting from the
equations (2a,b,c) are listed.
Since we restrict ourselves to irrotational flows, a {\sl LEMMA}
proved in BE93 applies, which enables us to reduce the
integrability conditions (2a,b,c) for the irrotationality of
the gravitational field strength to the conditions (BE93: 5d,e,f)
for the irrotationality of the velocity.
The components of the vorticity vector in Eulerian space
$\vec \omega = {1 \over 2} \nabla_x
\times \vec v$, $\vec v = \dot {\vec f}$, are expressed in Lagrangian
space and are listed up to the second order
for the ansatz (3) in (BE93, APPENDIX).
The additional third order terms resulting from the same ansatz read
explicitly (i,j,k = 1,2,3, cyclic ordering):
$$
\eqalignno{
\omega_i = \varepsilon^3 a \lbrack &
{\dot \psi}^{(2)}_{,i,k} {\psi}^{(1)}_{,i,j} -
{\dot \psi}^{(2)}_{,i,j} {\psi}^{(1)}_{,i,k} +
{\dot \psi}^{(2)}_{,j,k} {\psi}^{(1)}_{,j,j} -
{\dot \psi}^{(2)}_{,j,j} {\psi}^{(1)}_{,j,k} +
{\dot \psi}^{(2)}_{,k,k} {\psi}^{(1)}_{,k,j} -
{\dot \psi}^{(2)}_{,k,j} {\psi}^{(1)}_{,k,k} \cr &
{\dot \psi}^{(1)}_{,i,k} {\psi}^{(2)}_{,i,j} -
{\dot \psi}^{(1)}_{,i,j} {\psi}^{(2)}_{,i,k} +
{\dot \psi}^{(1)}_{,j,k} {\psi}^{(2)}_{,j,j} -
{\dot \psi}^{(1)}_{,j,j} {\psi}^{(2)}_{,j,k} +
{\dot \psi}^{(1)}_{,k,k} {\psi}^{(2)}_{,k,j} -
{\dot \psi}^{(1)}_{,k,j} {\psi}^{(2)}_{,k,k} \rbrack \cr
+ \varepsilon^3 \lbrack &
{\dot \psi}^{(1)}_{,i,j} ({\psi}^{(1)}_{,i,j}{\psi}^{(1)}_{,j,k} -
{\psi}^{(1)}_{,i,k}{\psi}^{(1)}_{,j,j})
+ {\dot \psi}^{(1)}_{,i,k} ({\psi}^{(1)}_{,i,j}{\psi}^{(1)}_{,k,k} -
{\psi}^{(1)}_{,i,k}{\psi}^{(1)}_{,j,k})
+ {\dot \psi}^{(1)}_{,j,k} ({\psi}^{(1)}_{,i,k}{\psi}^{(1)}_{,i,k} -
{\psi}^{(1)}_{,i,j}{\psi}^{(1)}_{,i,j}) \cr
+ & {\dot \psi}^{(1)}_{,j,k} ({\psi}^{(1)}_{,i,i}{\psi}^{(1)}_{,j,j} -
{\psi}^{(1)}_{,i,i}{\psi}^{(1)}_{,k,k})
+ {\dot \psi}^{(1)}_{,j,j} ({\psi}^{(1)}_{,i,j}{\psi}^{(1)}_{,i,k} -
{\psi}^{(1)}_{,i,i}{\psi}^{(1)}_{,j,k}) +
{\dot \psi}^{(1)}_{,k,k} ({\psi}^{(1)}_{,i,i}{\psi}^{(1)}_{,i,k} -
{\psi}^{(1)}_{,i,j}{\psi}^{(1)}_{,i,k}) \rbrack \,.\cr}
$$
The second order terms do not imply restrictions in the case of
solutions of the form (6), since a functional relationship between
the peculiar-acceleration and the peculiar-velocity
is assumed (see BE93). The third order terms depend on
the solutions (6a) and (6b). We insert them into the constraint
equations and obtain:
$$
\omega_i =\varepsilon^3 a ({\dot q}_{zz} q_z - {\dot q}_z q_{zz})\lbrack
{\cal S}^{(2)}_{,i,k} {\cal S}^{(1)}_{,i,j} -
{\cal S}^{(2)}_{,i,j} {\cal S}^{(1)}_{,i,k} +
{\cal S}^{(2)}_{,j,k} {\cal S}^{(1)}_{,j,j} -
{\cal S}^{(2)}_{,j,j} {\cal S}^{(1)}_{,j,k} +
{\cal S}^{(2)}_{,k,k} {\cal S}^{(1)}_{,k,j} -
{\cal S}^{(2)}_{,k,j} {\cal S}^{(1)}_{,k,k} \rbrack \;.
$$
The terms which are of third order in the first order perturbations
cancel for the present ansatz. We are left with constraints
which result from interaction terms of first and second
order perturbations. We conclude that interaction of the
perturbations add transversality to the third order solution which is
not covered by the longitudinal
ansatz. The Wronskian ${\dot q}_{zz} q_z -
{\dot q}_z q_{zz}$ is non-zero, because $q_z (t)$ and $q_{zz} (t)$
are linearly independent solutions. Thus,
in order to fulfill the constraint equations for the longitudinal
ansatz,
we have to assure a functional relationship of the form
$\nabla_0 {\cal S}^{(2)} = {\cal W} (\nabla_0 {\cal S}^{(1)})$
with arbitrary $\cal W$.
Since $\Delta_0 {\cal S}^{(2)} = 2 II({\cal S}_{,i,k})$, we
obtain the following equation for the functional $\cal W$
(a prime, here, denotes derivative with respect to the argument
$\nabla_0 {\cal S}^{(1)}$; without loss of generality we can set
${\cal S}^{(1)} = {\cal S}t_0$, see APPENDIX B):
$$
{\cal W}' I({\cal S}_{,i,k})
- 2 II({\cal S}_{,i,k}) t_0\;=\;0\;\;\;.
$$
Neglecting interaction terms altogether, the
third order terms are not constrained. The constraints are removed
by the transverse part derived in section {\bf 3.5.}.

}

\vfill\eject

\centerline{\rmm APPENDIX B}
\bigskip\bigskip
{\rmb
\noindent
In this APPENDIX a short discussion is given of
the uniqueness of the solutions
obtained. In particular, I shall demonstrate this
in the case of the
first order solution, and for the transverse part of the third order
solution.
\smallskip
\noindent
For the realization of the first order solution we replace the
perturbation potential ${\cal S}^{(1)}$ by the initial peculiar-velocity
potential ${\cal S}t_0$.
With this replacement the first order part of the
solution (24) reduces to the ``Zel'dovich approximation". This can be
done without loss of generality as will be shown below:

\noindent
The solution to equation (24f) can be written as
$$
{\cal S}^{(1)} = {\cal S}t_0 + \psi
$$
for any function $\psi$ which obeys the Laplace equation $\Delta_0 \psi
= 0$. Thus, if we preserve
the boundary conditions used for the
initial condition $\cal S$ (here assumed to
be periodic), then the only solution
of the Laplace equation is $\psi = \ell (t)$. Without loss of generality
we can set $\ell (t) =0$.
According to the form of the
solution (24), this argument holds for any time. $\;\;$ {\bf q.e.d.}

\bigskip
\noindent
For the transverse part of the third order solution a vector potential
$\vec {\cal S}^{(3c)}$ has been introduced (eq. (19a)). On this
a gauge condition has been imposed (eq. (19b)), which enables
to write the transverse third order solution (eqs. (24j,k,l)) in terms
of three Poisson equations. This is especially useful for technical
reasons to realize the solution ({\bf 4.3}). In the following it is
shown that the condition (19b) does not restrict the generality
of the solution:

\noindent
Firstly, note that the
vector $\vec \Xi = - \nabla_0 \times \vec {\cal S}^{(3c)}$ remains
transverse, i.e., $\nabla_0 \cdot \vec \Xi = 0$, if we add the gradient
of an arbitrary scalar function $\Omega$ to the vector potential
$\vec {\cal S}^{(3c)}$. The divergence of this vector potential is yet
unspecified. If we impose the gauge condition (19b), then
we have to choose the gauge function $\Omega$ in such a way that
$$
\Delta_0 \Omega = - \nabla_0 \cdot \vec {\cal S}^{(3c)} \;\;\;\;.
$$
This is always possible according to a theorem by Brelot on the
existence of solutions of Poisson equations (see Friedman 1963).
Imposing the gauge condition (19b) is sufficient to solve the equations
(24j,k,l), since (19b) implies $\nabla_0 (\nabla_0 \cdot \vec
{\cal S}^{(3c)}) = \vec 0$ required for the derivation of these
equations. $\;\;$ {\bf q.e.d.}
\bigskip\noindent
A comprehensive discussion of the problem of uniqueness in
general perturbation solutions will be given in a separate
paper (Ehlers \& Buchert 1993).

}

\vfill\eject

\centerline{\rmm References}
\bigskip\bigskip
{\rmb
Arnol'd V.I., Shandarin S.F., Zel'dovich Ya.B. (1982): {\it Geophys.
Astrophys.

Fluid Dyn.} {\bf 20}, 111.
\smallskip
Berezinsky V.S., Gurevich A.V., Zybin K.P. (1992): {\it Phys. Lett.}
{\bf B294}, 221.
\smallskip
Bertschinger E. (1985): {\it Ap.J.Suppl.} {\bf 58}, 39.
\smallskip
Bertschinger E. (1992): in: {\it New Insight into the Universe}, eds.:

V.J. Mart\'\i nez \etal,
pp. 65-126.
\smallskip
Bildhauer S., Buchert T. (1991): {\it Prog. Theor. Phys.} {\bf 86}, 653.
\smallskip
Bouchet F.R., Juszkiewicz R., Colombi S., Pellat R. (1992):
{\it Ap.J. Lett.} {\bf 394},

L5.
\smallskip
Buchert T., G\"otz G. (1987): {\it J. Math. Phys.} {\bf 28}, 2714.
\smallskip
Buchert T. (1989): {\it Astron. Astrophys.} {\bf 223}, 9.
\smallskip
Buchert T., Bartelmann M. (1991): {\it Astron. Astrophys.} {\bf 251},
389.
\smallskip
Buchert T. (1992): {\it M.N.R.A.S.} {\bf 254}, 729.
\smallskip
Buchert T. (1993): {\it Astron. Astrophys. Lett.} {\bf 267}, L51.
\smallskip
Buchert T., Ehlers J. (1993): {\it M.N.R.A.S.} {\bf 264}, 375.
\smallskip
Buchert T., Melott A.L., Wei\ss\ A.G. (1993a): {\it Astron. Astrophys.},
submitted.
\smallskip
Buchert T., Melott A.L., Wei\ss\ A.G. (1993b): {\it M.N.R.A.S.},

to be submitted.
\smallskip
Buchert T., Karakatsanis G., Klaffl R., Schiller P. (1993c):
{\it Astron. Astrophys.},

to be submitted.
\smallskip
Chernin A.D: (1993): {\it Astron. Astrophys.} {\bf 267}, 315.
\smallskip
Coles P., Melott A.L., Shandarin S.F. (1993): {\it M.N.R.A.S.}
{\bf 260}, 765.
\smallskip
Doroshkevich A.G. (1973): {\it Astrophys. Lett.} {\bf 14}, 11.
\smallskip
Doroshkevich A.G., Kotok E.V., Novikov I.D., Polyudov A.N.,

Shandarin S.F., Sigov Yu.S.
(1980): {\it M.N.R.A.S.} {\bf 192}, 321.
\smallskip
Ehlers J., Buchert T. (1993): {\sl in preparation}
\smallskip
Filmore J.A., Goldreich P. (1984): {\it Ap.J.} {\bf 281}, 1.
\smallskip
Friedman A. (1963): {\it Generalized functions and partial differential
equations,}

Englwood Cliffs, N.J.: Prentice Hall Inc., 1963.
\smallskip
Gurevich A.V., Zybin K.P. (1988a): {\it Sov. Phys. JETP} {\bf 67}, 1.
\smallskip
Gurevich A.V., Zybin K.P. (1988b): {\it Sov. Phys. JETP} {\bf 67}, 1957.
\smallskip
Gurevich A.V., Zybin K.P. (1990): {\it Sov. Phys. JETP} {\bf 70}, 10.
\smallskip
Jones B.J.T. (1993): in: {\it The Dennis Sciama $65^{th}$ Birthday
Celebration},

Cambridge University Press, in press.
\smallskip
Lachi\`eze-Rey M. (1993): {\it Ap.J.} {\bf 408}, 403.
\smallskip
Mandelbrot B. (1976): {\it Comptes Rendus (Paris)} {\bf 282A}, 119 (in
French).
\smallskip
Mart\'\i nez V.J. (1991): in: {\it Applying Fractals to Astronomy},
Lecture Notes in

Physics,
eds.: A. Heck, J. Perdang, Springer, pp. 135-159.
\smallskip
Melott A.L., Shandarin S.F. (1990): {\it Nature} {\bf 346}, 633.
\smallskip
Moutarde F., Alimi J.-M., Bouchet F.R., Pellat R., Ramani A.
(1991):

{\it Ap.J.} {\bf 382}, 377.
\smallskip
Nusser A., Dekel A., Bertschinger E., Blumenthal G.R. (1991):
{\it Ap.J.} {\bf 379}, 6.
\smallskip
Peebles P.J.E. (1980): {\it The Large-scale Structure of the Universe},

Princeton Univ. Press.
\smallskip
Peebles P.J.E. (1990): {\it Ap.J.} {\bf 365}, 27.
\smallskip
Zel'dovich Ya.B. (1970): {\it Astron. Astrophys.} {\bf 5}, 84.
\smallskip
Zel'dovich Ya.B. (1973): {\it Astrophysics} {\bf 6}, 164.

}

\vfill\eject

\bye